\newif\ifsubmodeapjs
\newif\ifsubmodeastroph

\submodeapjsfalse	
\submodeastrophtrue	

\ifsubmodeapjs
  \documentclass[12pt,preprint]{aastex}
  \usepackage{natbib,amsmath,verbatim}
  \received{}
  \revised{}
  \accepted{}
\fi
\ifsubmodeastroph
  \documentclass[12pt,preprint,iop]{emulateapj}
  \usepackage{natbib,amsmath,verbatim}

\fi

\newcommand{\etal}{{et al.~}}

\def\arcsec{\ensuremath{^{\prime\prime}}}

\def\etal{{\it et al.}}
\def\lae{\mathrel{<\kern-1.0em\lower0.9ex\hbox{$\sim$}}}
\def\gae{\mathrel{>\kern-1.0em\lower0.9ex\hbox{$\sim$}}}
\def\deg{^{\circ}}

\def\HST{{\it HST}}

\newcommand{\note}[1]{}

\def\calacs{{\tt calacs}}
\def\calwf{{\tt calwf3}}
\def\drizzle{{\tt Drizzle}}

\def\mosaicdrizzle{{\tt MosaicDrizzle}}
\def\multidrizzle{{\tt MultiDrizzle}}

\shorttitle{The 2012 Hubble Ultra Deep Field (UDF12): Observational Overview}

\shortauthors{Koekemoer et al.}

\ifsubmodeastroph
  \submitted{Submitted to the Astrophysical Journal Supplement, Dec 2012}
\fi

\begin{document}

\title{The 2012 Hubble Ultra Deep Field (UDF12): Observational Overview}

\author{
Anton M. Koekemoer\altaffilmark{1},
Richard S Ellis\altaffilmark{2},
Ross J. McLure\altaffilmark{3},
James S. Dunlop\altaffilmark{3},
Brant E Robertson\altaffilmark{4},
Yoshiaki Ono\altaffilmark{5},
Matthew A. Schenker\altaffilmark{2},
Masami Ouchi\altaffilmark{5},
Rebecca A. A. Bowler\altaffilmark{3},
Alexander B. Rogers\altaffilmark{3},
Emma Curtis-Lake\altaffilmark{3},
Evan Schneider\altaffilmark{4},
Stephane Charlot\altaffilmark{6},
Daniel P. Stark\altaffilmark{4},
Steven R. Furlanetto\altaffilmark{7},
Michele Cirasuolo\altaffilmark{3,8},
V. Wild\altaffilmark{3,9},
T. Targett\altaffilmark{3}
}
\altaffiltext{1}{Space Telescope Science Institute, 3700 San Martin Dr., Baltimore, MD 21218}
\altaffiltext{2}{Department of Astrophysics, California Institute of Technology, MS 249-17, Pasadena, CA 91125}
\altaffiltext{3}{Institute for Astronomy, University of Edinburgh, Royal Observatory, Edinburgh EH9 3HJ, UK}
\altaffiltext{4}{Department of Astronomy and Steward Observatory, University of Arizona, Tucson AZ 85721}
\altaffiltext{5}{Institute for Cosmic Ray Research, University of Tokyo, Kashiwa City, Chiba 277-8582, Japan} 
\altaffiltext{6}{UPMC-CNRS, UMR7095, Institut d'Astrophysique de Paris, F-75014, Paris, France}
\altaffiltext{7}{Department of Physics \& Astronomy, University of California, Los Angeles CA 90095}
\altaffiltext{8}{UK Astronomy Technology Centre, Royal Observatory, Edinburgh EH9 3HJ, UK}
\altaffiltext{9}{School of Physics and Astronomy, University of St Andrews, North Haugh, St Andrews, KY16 9SS}

\begin{abstract}

We present the 2012 Hubble Ultra Deep Field campaign (UDF12), a large 128-orbit Cycle 19 \HST\ program aimed at extending previous WFC3/IR observations of the UDF by quadrupling the exposure time in the F105W filter, imaging in an additional F140W filter, and extending the F160W exposure time by 50\%. The principal scientific goal of this project is to determine whether galaxies reionized the universe; our observations are designed to provide a robust determination of the star formation density at $z$$\,\gtrsim\,$8, improve measurements of the ultraviolet continuum slope at $z$$\,\sim\,$7$\,-\,$8, facilitate the construction of new samples of $z$$\,\sim\,$9$\,-\,$10 candidates, and enable the detection of sources up to $z$$\,\sim\,$12. For this project we committed to combining these and other WFC3/IR imaging observations of the UDF area into a single homogeneous dataset, to provide the deepest near-infrared observations of the sky currently achievable. In this paper we present the observational overview of the project, motivated by its scientific goals, and describe the procedures used in reducing the data as well as the final products that are produced. We have used the most up up-to-date methods for calibrating and combining the images, in particular paying attention to correcting several instrumental effects. We release the full combined mosaics, comprising a single, unified set of mosaics of the UDF, providing the deepest near-infrared blank-field view of the universe obtained to date, reaching magnitudes as deep as AB$\,\sim\,$30 in the near-infrared, and yielding a legacy dataset on this field of lasting scientific value to the community.

\end{abstract}

\keywords{ 
Cosmology: observations ---
Galaxies: high-redshift --- 
}

%
%
%

%

\section {Introduction}\label{section:intro}

A fundamental quest of modern observational cosmology involves expanding the frontiers of knowledge about the formation of the first stars and galaxies at the earliest epochs of cosmic time, and determining their role in the reionization of the universe at redshifts above 7. This is also among the most challenging of observational regimes to explore, requiring depths up to 30th magnitude (AB) or beyond, at count-rates that are thousands to millions of times fainter than the typical ground-based sky brightness per square arcsecond at optical and near-infrared wavelengths, respectively.

The {\it Hubble Space Telescope (HST)} has played a unique role in these explorations of the very early universe, probing these extreme depths by virtue of its combination of high angular resolution, and low sky background achievable only from space. Significant investments of time in deep single-pointing surveys with \HST\ have yielded a wealth of scientific results to date, from the original Hubble Deep Field
	\citep{1996AJ....112.1335W}\note{Williams \etal 1996}
and Hubble Deep Field South
	(\citealt{2000AJ....120.2747C}\note{Casertano et al. 2000};
	\citealt{2000AJ....120.2735W}\note{Williams et al. 2000};
	\citealt{2003AJ....125..398L}\note{Lucas et al. 2003}),
together with the 2004 Ultra Deep Field (UDF:
	\citealt{2006AJ....132.1729B}\note{Beckwith et al. 2006};
	\citealt{2005AJ....130....1T}\note{Thompson et al. 2005}),
which has since become the centerpoint for deep follow-on imaging programs in 2005 (PI.: M. Stiavelli, described in
	\citealt{2007ApJ...671.1212O,2009ApJ...690.1350O}\note{Oesch et al. 2007,2009}),
as well as with the Wide Field Camera 3 infrared channel (WFC3/IR) in 2009 (UDF09, PI. G. Illingworth, described in
	\citealt{2010ApJ...709L..16O,2010ApJ...709L..21O}\note{Oesch et al. 2010a,b}
	and
	\citealt{2011ApJ...737...90B}\note{Bouwens et al. 2011})
and most recently in 2012 (UDF12, PI: R. Ellis, described in	
	\citealt{2012arXiv1206.0737E}\note{Ellis et al. 2012}
together with the present paper).%
	\footnote{Note that there is a separate UDF imaging program, in the ultraviolet with WFC3/UVIS (PI.: H. Teplitz); this probes lower redshifts complementary to the very distant universe discussed here.}

\begin{deluxetable*}{lllccc}
\tablecaption{\label{table:udf12obs}
	UDF12 Observing Summary (HST Program ID 12498)}
\tablehead{%
Field		& Instrument/Camera	& Filter	& \# Orbits	& \# Exposures	& Exposure Time (s)}
\startdata
UDF-main	& WFC3/IR		& F105W		& 72		& 144		& 198,423	\\
		& WFC3/IR		& F140W		& 30		& 60		& 82,676	\\
\vspace{1mm}
		& WFC3/IR		& F160W		& 26		& 52		& 71,652	\\
UDF-par2	& ACS/WFC		& F814W		& 128		& 256		& 322,944
\enddata
\hfill\break
\end{deluxetable*}

These surveys, in conjunction with wider, shallower \HST\ surveys with the Advanced Camera for Surveys (ACS) including GOODS
	\citep{2004ApJ...600L..93G}\note{Giavalisco et al. 2004},
GEMS
	\citep{2004ApJS..152..163R}\note{Rix et al. 2004},
AEGIS
	\citep{2007ApJ...660L...1D}\note{Davis et al. 2007},
COSMOS
	\citep{2007ApJS..172....1S,2007ApJS..172..196K}\note{Scoville et al. 2007; Koekemoer et al. 2007},
WFC3 Early Release Science program
	\citep{2011ApJS..193...27W}\note{Windhorst et al. 2011},
CANDELS
	\citep{2011ApJS..197...35G,2011ApJS..197...36K}\note{Grogin et al. 2011; Koekemoer et al. 2011},
BoRG
	\citep{2011ApJ...727L..39T}\note{Trenti et al. 2011},
HIPPIES
	\citep{2011ApJ...728L..22Y}\note{Yan et al. 2011a},
and CLASH
	\citep{2012ApJS..199...25P}\note{Postman et al. 2012},
have transformed our understanding of the early universe. There is now overall evidence for the mass build-up of early galaxies at $z$$\,\sim\,$4$\,-\,$8 based on the evolution of the cosmic star-formation density
	(\citealt{2004ApJ...600L..93G}\note{Giavalisco et al. 2004b},
	\citealt{2004ApJ...616L..79B,2007ApJ...670..928B}\note{Bouwens et al. 2004, 2007},
	\citealt{2004MNRAS.355..374B}\note{Bunker et al. 2004},
	\citealt{2006MNRAS.372..357M,2009MNRAS.395.2196M}\note{McLure et al. 2006, 2009},
	\citealt{2006ApJ...651...24Y,2010RAA....10..867Y}\note{Yan et al. 2006, 2010},
	\citealt{2010A&A...511A..20C}\note{Castellano et al. 2010},
	\citealt{2010ApJ...709L..16O}\note{Oesch et al. 2010a}).
A wider variety of results have been obtained on the ultraviolet spectral slopes and stellar populations of these early star-forming galaxies at $z$$\,\sim\,$7$\,-\,$8
	(\citealt{2009ApJ...705..936B,2010ApJ...708L..69B,2012ApJ...754...83B}\note{Bouwens et al. 2009, 2010c, 2012},
	\citealt{2010ApJ...724.1524O}\note{Ono et al. 2010},
	\citealt{2010MNRAS.409..855B}\note{Bunker et al. 2010},
	\citealt{2010ApJ...719.1250F,2012ApJ...756..164F,2012ApJ...758...93F}\note{Finkelstein et al. 2010, 2012a,b},
	\citealt{2011ApJ...728L..22Y,2011arXiv1112.6406Y}\note{Yan et al. 2011a,b},
	\citealt{2010MNRAS.403..960M,2011MNRAS.418.2074M}\note{Mclure et al. 2010, 2011},
	\citealt{2011A&A...532A..33G,2012A&A...547A..51G}\note{Grazian et al. 2011, 2012},
	\citealt{2012ApJ...760..108B}\note{Bradley et al. 2012},
	\citealt{2012MNRAS.420..901D,2012arXiv1212.0860D}\note{Dunlop et al. 2012}),
particularly concerning the extent to which these galaxies may or may not be able to account for reionization. Finally, tantalizing discoveries of galaxies at $z$$\,\sim\,$9$\,-\,$10
	(\citealt{2010RAA....10..867Y}\note{Yan et al. 2010},
	\citealt{2011Natur.469..504B}\note{Bouwens et al. 2011b},
	\citealt{2012Natur.489..406Z}\note{Zheng et al. 2012}),
$z$$\,\sim\,$11
	\citep{2012arXiv1211.3663C}\note{Coe et al. 2012},
and most recently up to $z\,\sim\,$12
	\citep{2012arXiv1206.0737E}\note{Ellis et al. 2012}
are only becoming possible by means of the deepest near-IR observations achievable, which for the UDF are described in this paper.

The structure of this paper is as follows. The survey overview and observational design are presented in \S\ref{section:observations}, followed by the description of the data processing and calibration in \S\ref{section:processing}, the presentation of the final data products in \S\ref{section:data}, and the summary in \S\ref{section:summary}. Further details and current updates about the survey are provided at the project website%
	\footnote{http://udf12.arizona.edu/}
while all the final combined mosaic data products from the survey are being made publicly available as High-Level Science Products%
	\footnote{http://archive.stsci.edu/prepds/hudf12/}
that are delivered to the Space Telescope Science Institute archive. These images constitute a single, unified set of mosaics of the UDF, providing the deepest near-IR blank-field view obtained of the universe to date, approaching limiting magnitudes AB$\,\sim\,$30 in the near-IR, and yielding a legacy dataset on this field that is of lasting scientific value.

\section{Survey Overview and Observational Design}\label{section:observations}

This paper presents the overview of the 2012 Hubble Ultra Deep Field campaign (UDF12, \HST\ Program ID 12498, PI.: R. Ellis), a large 128-orbit Cycle 19 \HST\ program aimed at extending the previous 2009 WFC3/IR observations of the UDF (UDF09, \HST\ Program ID 11563, PI.: G. Illingworth). The observations were all obtained between 4 August 2012 and 16 September 2012, and are summarized in Table~\ref{table:udf12obs}. The observational approach as proposed by the current project is to combine these and other WFC3/IR imaging observations of the UDF area into a single homogeneous dataset, including additional filter wavelength coverage, to provide the deepest near-IR observations of the sky currently achievable, as summarized in Table~\ref{table:fulldepth}.

\begin{deluxetable}{lcccc}
\tablecaption{\label{table:fulldepth}
	Full-Depth Combined UDF WFC3/IR mosaics (orbits)}
\tablehead{%
Filter		& UDF09\tablenotemark{a}
				& UDF12\tablenotemark{a}
						& Other\tablenotemark{a,b}
								& Final\tablenotemark{a}}
\startdata
F105W		& 24		& 72		& 4		& 10	\\
F125W		& 34		& $-$		& 5		& 39	\\
F140W		& $-$		& 30		& $-$		& 30	\\
F160W		& 53		& 26		& 5		& 84
\enddata
\tablenotetext{a}{Number of HST orbits obtained.}
\tablenotetext{b}{Other programs include 12099 (PI: A. Riess) and CANDELS 12060,12061,12062 (PI: S. Faber, H. Ferguson).\hfill\break\hfill\break}
\end{deluxetable}

In this project, we aim to study the role of galaxies in reionizing the universe, by extending robust searches for Lyman-break galaxies to $z$$\,\sim\,$9 and beyond, obtaining more accurate faint-end luminosity functions at z$\,\sim\,$7 and z$\,\sim\,$8, and determining more accurate ultraviolet spectral energy distributions to constrain stellar populations and ionizing photon output. In order to achieve these goals, this survey builds upon the previous WFC3/IR investment that had already been obtained in the UDF, in a number of different ways:
\begin{itemize}
\item Quadruple the exposure time in the F105W filter, adding 72 new orbits to the data that had previously been obtained in the UDF09 survey (program ID 11563, PI.: G. Illingworth), in order to provide deeper short-wavelength constraints on the $z$$\,\sim\,$8 sources selected using F105W$\,-\,$F125W color criteria, probe to fainter luminosities, and yield a more robust determination of the star formation density at  $z$$\,\sim\,$8$\,-\,$10.
\item Add completely new wavelength information with the F140W filter, obtaining 30 orbits of deep integration to match the depths in the F125W and F160W filters. This provides improved measurements of the UV slopes of $z$$\,\sim\,$7$\,-\,$8 sources, additional independent detections of the continuum longward of the Lyman break for sources at $z$$\,\gtrsim\,$9$\,-\,$10, as well as probing to $z$$\,\sim\,$12 for sources whose Lyman emission may be redshifted out of the F140W filter and are detectable only in F160W.
\item Increasing the exposure time in F160W by an additional 26 orbits from the UDF09 program, providing more robust red measurements for galaxies at $z$$\,\sim\,$8$\,-\,$10 and further improving the constraints on their UV slopes, as well as further securing any potential detections up to $z$$\,\sim\,$12.
\end{itemize}
In this section we provide the details of the observational design of the survey, including the filter selection, exposure times, \HST\ observing considerations, and dithering strategies that were employed in obtaining the observations.

\subsection{Filter Wavelength Coverage and Depth}

The science drivers, as summarized above and described here in further detail, drove the filter choice and exposure times at a high level to achieve 5$\sigma$ limiting depths of F105W$\,=\,$30.0, F140W$\,=\,$29.5, and F160W$\,=\,$29.5 (AB magnitudes, as measured in 0$\farcs$4 diameter apertures), in conjunction with F125W$\,=\,$29.5 from reprocessing the previously obtained observations.

The increase in F105W depth was necessitated by the need to improve the robustness of the $z$$\,\sim\,$8 dropout selection as well as probing to fainter luminosities. The lack of dynamic range offered in earlier data between the F105W$\,-\,$F125W color selections, corresponding only to a 2$\sigma$ limit of F105W$\,-\,$F125W$\,>\,$1.0 at the F125W detection limits, was inadequate to exclude potential low-redshift Balmer-break interlopers. Increasing this discriminating power to a 2$\sigma$ limit of F105W$\,-\,$F125W$\,>\,$1.5, at a 5$\sigma$ detection threshold of F125W$\,=\,$29.5 obtained from our own reduction of the previous UDF09 data (program ID 11563, PI.: G. Illingworth), provides much stronger selection against low-redshift interlopers, hence cleaner samples of $z$$\,\gtrsim\,$8 galaxies for studies of the luminosity function evolution at these redshifts. We achieved this increase with the additional 72 orbits that we obtained, thereby reaching our target 5$\sigma$ detection limit of F105W$\,=\,$30.0 (AB magnitudes, 0$\farcs$4 diameter aperture).

The depth increase for F160W, aiming to match the depth of our F125W reduction, was motivated primarily by the need to probe further down the luminosity function at redshift $z$$\,\simeq\,$7 and, for the first time, at $z$$\,\simeq\,$8. In particular, establishing whether or not the galaxy population at $z$$\,\sim\,$8 can reionize the universe requires measuring the faint end slope of the luminosity function down to $M_{\rm 1500}$$\,\sim\,$$-$17.5, thereby necessitating 5$\sigma$ detection limits of $\sim\,$29.5 (AB). This was achieved through increasing the existing F160W exposure time by an additional 50\% to reach a 5$\sigma$ detection limit of F160W$\,=\,$29.5, from combining our new 26 orbits with the previously existing F160W data on this field.

Finally, the addition of the new F140W filter was motivated by several science goals. First among these was to improve the reliability of any detections of possible sources at $z$$\,\sim\,$9$\,-\,$10 (whose Lyman break has moved out of the F125W filter) by using two filters to provide detections longward of the break. Since  high-redshift galaxies are essentially flat-spectrum sources in $f_\nu$, this then drives the depth of F140W to match that of F160W, i.e., a 5$\sigma$ detection threshold of F140W$\,=\,$29.5 (AB), which we achieved using a total exposure time of 30 orbits in this filter. This filter also provides additional wavelength discriminating power for sources that might be at even higher redshift, since Lyman-$\alpha$ would move out of its redward edge at $z$$\,\sim\,$12.

The additional depth in the F105W and F160W filters, along with the new wavelength information provided by the F140W filter to a matching depth, also provide a more accurate measurement of the UV slope parameter $\beta$; a robust measure of the average value of $langle$$\beta$$rangle$ requires 8$\sigma$ detections to achieve reliable constraints. For example, as demonstrated in
	Dunlop et al. (2012),		
shallower 4$\sigma$ detections can lead to colors which may be uncertain by up to $\sim\,$0.35 magnitudes, translating to 1$\sigma$ errors in $\beta$ of $\Delta$$\beta$$\,\pm\,$1.5. Doubling the S/N on the measurements of $\beta$, obtained by using all the new data from our program, can therefore estalish much more accurate values of  $langle$$\beta$$\rangle$ for galaxies as faint as $M_{UV}$$\,\sim\,$$-$18 at $z$$\,\sim\,$7. In addition, the increased depth, along with the new F140W information, can extend the measured values of $\beta$ for galaxies up to at least $z$$\,\sim\,$8.

In addition to the prime WFC3/IR observations on the UDF, we also obtained parallel observations with the ACS/WFC camera. We required an orient designed to place all these parallel exposures onto the existing parallel field 2 of the UDF (labelled as UDF-PAR2, also known previously as UDF-NICP34), using a spacecraft orienation of 264$\deg$, identical to that which had been used for this field in the UDF09 program, to ensure maximal overlap with the ACS data that had been obtained as part of that program. As discussed further by
	McLure et al. (2011),		
the existing optical ACS data in that field was insufficient to properly exploit the new WFC3/IR imaging for the selection of galaxies at $z$$\,\gtrsim\,$6.5, and devoting this time to accumulating deep F814W imaging for all the parallel exposures was chosen as the most efficient way of improving this situation for maximum legacy value.

\begin{deluxetable}{llc}
\tablecaption{\label{table:zeropoints}
	WFC3/IR and ACS/WFC Zeropoints\tablenotemark{a}}
\tablehead{%
Instrument/Camera		& Filter			& Zeropoint (ABmag)}
\startdata
WFC3/IR				& F105W				& 26.269	\\
WFC3/IR				& F125W\tablenotemark{b}	& 26.230	\\
WFC3/IR				& F140W				& 26.452	\\
WFC3/IR				& F160W				& 25.946	\\
ACS/WFC\tablenotemark{c}	& F814W				& 25.947
\enddata
\tablenotetext{a}{Current information on zeropoints is available at:\hfill\break
	http://www.stsci.edu/hst/acs/analysis/zeropoints\hfill\break
	http://www.stsci.edu/hst/wfc3/phot\_zp\_lbn}
\tablenotetext{b}{In this paper we present information for F125W, for which we did not obtain new observations but instead provide an improved reprocessing of existing data, for consistency with the other observations that we obtained.}
\tablenotetext{c}{Our ACS/WFC data are obtained as parallel exposures on the UDF-par2 field, and do not overlap our WFC3/IR observations obtained on the main UDF field.\hfill\break}
\end{deluxetable}

We list in Table~\ref{table:zeropoints} the values of the zeropoints corresponding to the four WFC3/IR bandpasses that we used. We also indicate locations where more updated information may be available, if necessary. These zeropoints have an accuracy of at least $\sim\,$1$\,-\,$2\%; remaining uncertainties may be related to time-dependent changes in the filter or instrument properties, or improved knowledge of the standard stars that are used in determining the calibrations.

\subsection{HST Observations and Dither Patterns}

Due to HST scheduling constraints on our UDF12 program, we divided the 128 orbits into a total of 64 visits, where each visit consisted of two orbits, and each orbit consisted of two prime WFC3/IR exposures, accompanied by two parallel exposures using the ACS/WFC camera. We adopted similar observing strategies to previous programs in order to provide uniformity within the final combined datasets. In particular, our dither patterns followed a strategy consistent with that of the previous UDF09 observations
	(Oesch et al. 2010;
	Bouwens et al. 2011),
which in turn had been based on the original UDF dither strategy
	(Beckwith et al. 2006).
Each of the 64 visits (being 4 exposures each) consisted of a 4-point dither pattern, shifted around the sub-pixel phase space to provide the best possible PSF sub-sampling on both the WFC3/IR detector with its relatively large pixel scale (0$\farcs$128\,pixel$^{-1}$), as well as on the ACS/WFC CCDs (0$\farcs05$\,pixel$^{-1}$). In addition, each of these 64 sets of 4-point dithers was in turn offset onto a larger-scale grid, with offsets of $\sim\,\pm$3\,arcsec to cover the ACS/WFC chip gap and mitigate large defects and persistence on the WFC3/IR detector, as well as providing additional sub-pixel phase space sampling.

All our near-infrared UDF12 observations were obtained with the WFC3/IR detector, using the IR-FIX aperture which samples the full imaging field of view (1014$\times$1014 pixels, covering a region $\sim 130\arcsec$ across, with a plate scale of $0\farcs128$ pixel$^{-1}$ at its central reference pixel). All exposures were obtained in MULTIACCUM mode using the SPARS100 read-out sequence, enabling the pixels to be sampled nondestructively every 100$\,$s. These SPARS read samples were repeated for each sequence by specifying the NSAMP parameter, which was set to either 15 or 16 before reading out the array, corresponding to either 1300 or 1400 seconds, depending on the scheduling constraints for each particular orbit. Each of the MULTIACCUM sequences also included two short reads at the start, separated by 2.9$\,$s, which served to provide a measure of the bias structure across the array at the start of each exposure.

The parallel optical UDF12 observations were all obtained using the ACS/WFC camera, which comprises two CCDs with a usable area of $4096 \times 2048$ pixels, covering $\sim 200\arcsec$ in extent with a plate scale of $0\farcs05$ pixel$^{-1}$ at the central reference pixel.  The two detectors are located adjacent to one another with a small physical gap between them of $\sim 2\farcs5$. We used the WFC aperture for all the ACS exposures. The exposure times ranged between 1200$\,-\,$1300 seconds, depending upon orbital visibility constraints.

\section{Data Calibration and Processing}\label{section:processing}

Our final combined UDF09+UDF12 images mosaics have been processed with a version of the ``\mosaicdrizzle'' image combination pipeline, specially modified for the UDF12 program (see
	Koekemoer et al. 2002,		
        		2011		
for a more general description). This performs astrometric alignment and registration, cosmic ray rejection, and final combination of the exposures using the \multidrizzle\ software
	\citep{2002hstc.conf..337K}\note{Koekemoer et al. 2002},
as well as the \drizzle\ software
	\citep{2002PASP..114..144F}\note{Fruchter \& Hook 2002}.
In this section we provide descriptions of the input datasets, as well as the processing that was carried out within ``\mosaicdrizzle'', along with the resulting characteristics of the mosaics that were produced. We also processed all the UDF09 WFC3/IR data on the main UDF in a similar way, as well as all the other overlapping WFC3/IR data that were previously described.

\subsection{Initial WFC3/IR Standard Calibration}

We initially processed all our raw WFC3/IR images through standard calibration using the Pyraf/STSDAS task \calwf%
	\footnote{Further documentation for all the PyRAF/STSDAS data reduction software is provided at http://stsdas.stsci.edu/}
in order to obtain a first-pass set of calibrated images and carry out initial data quality validation. This task populates the bad pixel arrays using known bad pixel tables, and subtracts the bias for each read using the reference pixels around the border of the detector. It then carries out a subtraction of the zeroth read in order to remove the bias structure across the detector, followed by a subtraction of the dark current reference files for the SPARS100 read-out sequences. This was followed by the non-linearity correction and photometric keyword calculation, using the current filter throughput tables and detector quantum efficiency curves.

While the initial calibration was carried out using the standard pipeline dark reference files, we found that we could improve the signal-to-noise in the final mosaics by constructing a custom dark frame from the full set of dark calibration files that have been obtained on-orbit for the same readout mode and exposure times that we were using. Therefore we constructed such a dark frame and used it to recalibrate all the exposures, including our own as well as those from all the previous WFC3/IR observations on this field.

After having removed basic instrumental effects from each read, the exposures were then passed through the up-the-ramp slope fitting and cosmic ray rejection steps in \calwf. For each pixel, this step performs a linear fit to the accumulating counts that are sampled during each MULTIACCUM read, while rejecting outliers from the fit as being due to cosmic rays. A final count-rate value was then computed for each pixel using only the unflagged reads, and was stored as the count-rate in the final calibrated exposure, while the uncertainty in the slope of counts versus time was stored in the error extension of the image.

\begin{figure*}[t!]
\begin{center}
\ifsubmodeapjs
  \includegraphics[width=7in]{sky-background-corr.jpg.eps}
\fi
\ifsubmodeastroph
  \includegraphics[width=7in]{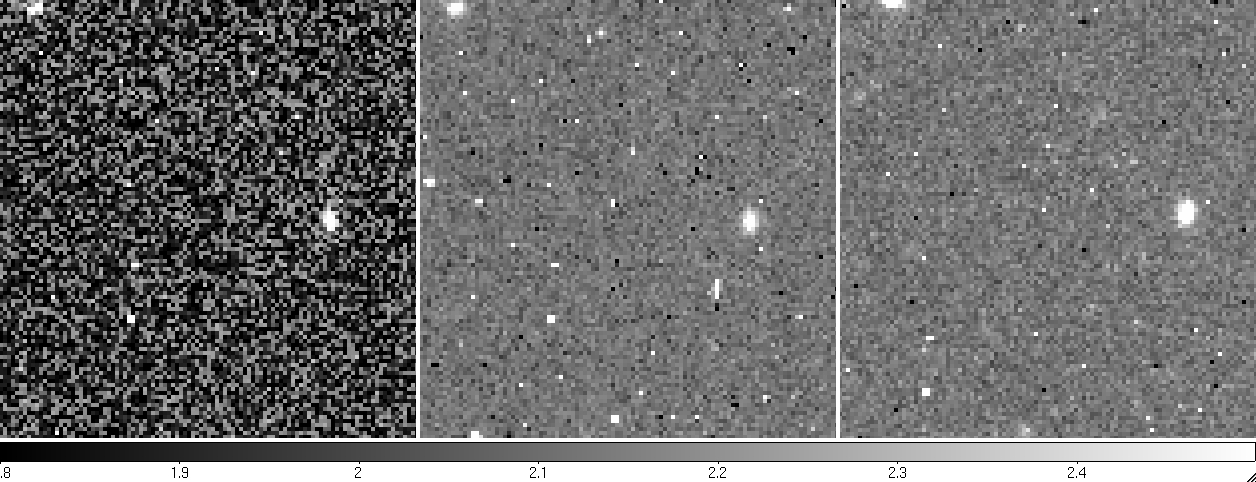}
\fi
\figcaption{\label{figure:variable_sky}%
Demonstration of the impact of time-dependent variable sky background on the WFC3/IR images, as well as the correction for this effect. All the images are shown to the same greyscale stretch. (Left) initial image after original calibration using the default \calwf, showing strongly non-Gaussian noise properties; (Middle) resulting image once intermediate correction has been applied for the time-dependent variable sky and subsequent recalibration, which we have developed for this project; (Right) similar exposure that was not affected by time-dependent variable sky; the noise properties of the corrected images (e.g., middle) are now similar to normal images such as those on the right, and can be safely combined in the mosaic without adversely impacting its final signal-to-noise.
}
\end{center}
\end{figure*}

\subsection{Additional Processing for WFC3/IR Exposures with Time-Dependent Variable Sky Background}

A fraction of the UDF12 exposures were affected by sky background emission that varied significantly as a function of time during the course of the read-out sequences. This was most noticeable in the F105W filter, due to the wavelength-dependent nature of the background sky emission. The resulting time-dependent sky background variation introduces a non-linear component into the counts that are measured at each read during the MULTIACCUM sequence. A consequence of this is that the standard \calwf\ up-the-ramp cosmic ray rejection and count-rate slope-fitting algorithm breaks down for these exposures, since that algorithm is designed for count-rates that are constant in time. Therefore, the resulting count-rate images after the completion of \calwf\ can have significantly non-Gaussian noise properties, which can adversely impact the signal-to-noise of the final mosaics, by up to several tenths of a magnitude or more, if they are included as-is.

To mitigate the impact of this issue on the final mosaics created for the UDF12 project, we developed a specialized set of routines to carry out intermediate processing of the partially calibrated MULTIACCUM sequences, after the initial bias and dark current subtraction had been performed by \calwf, but before running the count-rate slope-fitting. Specifically, these routines carried out a fit to any non-linear components that were present in the measured counts at each read, defined in such a way that the total integrated non-linear component was zero along the time sequence. Thus the non-linear component could be subtracted from these reads without any net impact on the total accumulated count-rates.

The corrected exposures were subsequently passed through the remaining steps in \calwf, particularly the up-the-ramp cosmic ray rejection and slope-fitting routines, which now performed successfully since the incoming counts were all linear in time. It was verified that in all cases, the resulting images had pixel distributions that were now perfectly Gaussian, reflecting the mean sky background as well as the correct corresponding noise properties. Furthermore, the photometry of sources extracted from these images was identical to that obtained from exposures that had no time-dependent sky background. Therefore, after passing all tests on their photometric and statistical properties, these images were subsequently included among all the others in the final mosaic combinations, enabling the required depths to be achieved.

\subsection{WFC3/IR Persistence, Warm Pixels and Flat Field Correction}

After the default calibrations had been applied, we implemented in our custom pipeline several further corrections to improve the WFC3/IR data, which are not part of the standard \calwf\ pipeline. The first of these concerns the presence of persistent flux in certain pixels due to bright sources having been observed in previous exposures, which can be a significant issue for the WFC3/IR detector.%
	\footnote{http://www.stsci.edu/hst/wfc3/ins\_performance/persistence/}
In some cases, we were able to make use of darks from the WFC3 calibration program that had executed just prior to our UDF12 visits, to aid in identifying and measuring problematic pixels. Pixels with persistent flux were then identified in these dark frames if they exceeded a count-rate threshold of five sigma above the mean, and were flagged in the following science exposures. For subsequent orbits during a visit, we could then determine directly from the preceding exposures which pixels may contain sufficient flux to cause persistence; the calibration darks were only needed for the first orbit in a visit, when the previous data may be from another program and not necessarily accessible. In those cases, we also made use of the persistence masks created for all exposures, accessible from the aforementioned website maintained by the WFC3 team, which we verified were successful in excluding all pixels that were affected by persistence.

We also identified additional ``warm'' pixels, using the full set of on-orbit dark exposures obtained during the UDF12 campaign to identify these pixels if they exceeded a threshold of five sigma above the mean, in which case they were flagged in the data quality arrays that were associated with each image, and were excluded from the final image combination. We further assembled median stacks of all exposures, flagging any pixels that varied significantly compared to the general population, which resulted in a small number of additional pixels being flagged that were not caught from the dark files.

Finally, the WFC3/IR detector is subject to IR ``blobs'' that have appeared in the WFC3/IR channel since launch and were not present in the ground flats, as well as some residual large-scale variation in the overall structure of the flatfield. While these are accounted for to some extent in the current calibration files, they remained noticeable in the deep combined UDF12 imaging. We therefore made sure to mask out all regions affected by the blobs in each exposure, as well as applying large-scale, low-level residual corrections to the flatfields as needed. We verified that the resulting images were flat to within $\sim\,$1$\,-\,$2\% of the mean sky level.

\subsection{ACS/WFC Calibration}

Each of the raw ACS/WFC exposures in the UDF12 parallel field were initially calibrated using the Pyraf/STSDAS task \calacs. This included bias subtraction, dark current correction, bad pixel masking and flatfielding. In addition, a number of other corrections need to be applied to ACS data, given the length of time that the instrument has been on orbit, as well as accounting for electronic effects in the new CCD Electronix Box Replacement (CEB-R) that was installed during Servicing Mission 4 (SM4). The first of these involves the correction for bias striping noise
	\citep{2010.Grogin.bias}
which is introduced by the electronics and manifests itself as a bias amplitude variation from one row to the next.

In addition, \calacs\ corrects for the impact of Charge Transfer Efficiency (CTE) degradation, whereby charge traps present in the pixels can capture some of the electrons during readout, leading to a loss of flux in the original pixel. This manifests itself as deferred-charge trails along the columns behind bright pixels in each exposure, while also producing a net astrometric shift up along in the column for bright sources. The effect becomes increasingly severe for pixels furthest from the amplifiers, which for these detectors are the pixels near the chip gaps. A key point about the CTE correction algorithm
	(\citealt{2010PASP..122.1035A}\note{Anderson \& Bedin 2010}).
is that it is effectively a deconvolution, by virtue of the fact that it restores the charge profiles of pixels along a particular column to their original shape, which is sharper and more concentrated than the observed profiles which have been smeared by the deferred charge trails. As such, the pixel-to-pixel noise in the final reconstructed image is also somewhat higher than in the original exposure. Tests to date have shown that this algorithm correctly reproduces the expected noise that would be present in the images if no CTE degradation had been present, and that it restores both the photometry and the astrometric accuracy to levels that are comparable to images without CTE degradation.

Finally, our UDF12 parallel ACS pipelines implement a routine to correct for additional bias-related offsets between the ACS/WFC3 detector amplifier quadrants, that are not fully corrected during standard calibration. This routine fits for the differences between quadrants, using an iterative clipping procedure to eliminate signal from astronomical sources and preserve only the background flux, which then removes the residual amplifier quadrant differences and places all four quadrants on a uniform background level.

\subsection{Relative Astrometry and Distortion}

Once the individual prime WFC3/IR UDF09+UDF12 exposures and parallel ACS/WFC exposures had all been passed through the initial calibrations at the detector level, they were subsequently passed through the rest of our astrometric and mosaicing processing pipelines. The first stage of this pipeline processes all the exposures in each visit, for the different cameras, and addresses the relative shifts between exposures in each given single-orbit visit. The accuracy of the astrometric information in the image headers depends on the pointing accuracy of the spacecraft, as well as the calibration of the geometric distortion models for the detectors. These are described in further detail for ACS/WFC in
	\citep{2007.Anderson.ISR}.	
For WFC3/IR we made use of distortion models published by
	(Kozhurina-Platais et al. 2012).		

\subsection{Cross-Correlation Shift Determination}

To further solve for and remove the residual uncertainties in the spacecraft dither offsets between all the exposures in each orbit, we applied cross-correlation procedures to all the exposures. The cross-correlation procedure first passes all the exposures through a partial run of \multidrizzle, up to the point where single-drizzled images are produced for all the individual exposures, which are all aligned on the same pixel grid so that astronomical sources should be at the same pixel locations if no residual shifts were present. These images were then masked, retaining only regions around objects that contain sufficient signal, and then passed through cross-correlation. The resulting cross-correlation peak was fit for each exposure using a two-dimensional fitting routine to determine its location and associated uncertainty, which was then directly translated into shifts. This procedure was iterated a few times for each exposure, in order to ensure convergence of the cross-correlation, with final uncertainties typically less than a few hundredths of a pixel

\subsection{Final Cosmic Ray and Bad Pixel Rejection}

After having corrected the relative shifts for all the exposures, final cosmic ray and bad pixel rejection was carried out for all the exposures of a given filter, for each camera, by carrying out another run of \multidrizzle, this time with the improved relative shifts. The cosmic rays were identified in the driz\_cr step of \multidrizzle\ using a process that first created a series of separately drizzled images, one for each input exposure, which were subsequently used to create a median image using the ``minmed'' algorithm in \multidrizzle, which enables the minimum to be used instead of the median in cases where valid pixels from only two or three exposures are present, if one or more of those are impacted by a cosmic ray.

The clean median image was then transformed back to the distorted detector frame of each input exposure to carry out cosmic ray rejection using the following approach. The input counts in a given pixel in the original exposure, $I_{\rm exp}$, were compared with the counts from the median image, $I_{\rm med}$, for the same pixel, together with the derivative of the median image, $\Delta_{\rm med}$, defined as the steepest gradient from that pixel to its surrounding pixels (with all these quantities being in units of electrons). A pixel was flagged as a cosmic ray if it exceeded a certain threshold depending on the background sky r.m.s. as well as the local gradient around that pixel.
The inclusion of the gradient term $\Delta_{\rm med}$ effectively ``softens'' the cosmic ray rejection in regions of relatively steep gradients such as bright cores of objects, where the pixel-to-pixel variation can exceed simple Poissonian statistics. For the data processing in our UDF12 pipeline, this rejection was performed over two iterations, with the first pass going through all the pixels in the image and using $S = 1.2$ and $SNR = 3.5$, followed by a second pass in a 1-pixel wide region around each of the pixels flagged in the first pass, but using more stringent criteria of $S = 0.7$ and $SNR = 3.0$. This ensured that fainter pixels around cosmic rays are also flagged.

\subsection{Absolute Astrometry}\label{section:absolute_astrometry}

We obtained absolute astrometry for all the visits relative to one another by using the original UDF catalogs
	(Beckwith et al. 2006),
as well as our own catalog that was generated on the UDF-PAR2 field from the existing mosaics
	(Bouwens et al. 2011).
All the sources detected in the exposures for each orbit were matched to the sources in the relevant portion of the absolute astrometric catalogs, using a number of iterative steps. The first iteration uses a relatively large tolerance (up to a few arcseconds) and only the brightest $\sim\,$20$\,-\,$30 sources in each image, in order to determine the dominant terms in the shifts for right ascension and declination. Once these had been accounted for, several additional iterations were carried out using the full catalog of sources in each image, using progressively tighter matching tolerances down to 0$\farcs$1 and solving for the residual remaining shifts as well as the rotation errors due to the uncertainties in guidestar position. For all visits, $\sim \,$300$\,-\,$400 sources were typically matched at the faintest levels and tightest tolerances between the {\it HST} \multidrizzle-combined images and the reference catalogs.

\begin{figure*}[h]
\begin{center}
\ifsubmodeapjs
  \includegraphics[width=3in]{udf_matched_cat_radec_dra_ddec.eps}
\fi
\ifsubmodeastroph
  \includegraphics[width=3in]{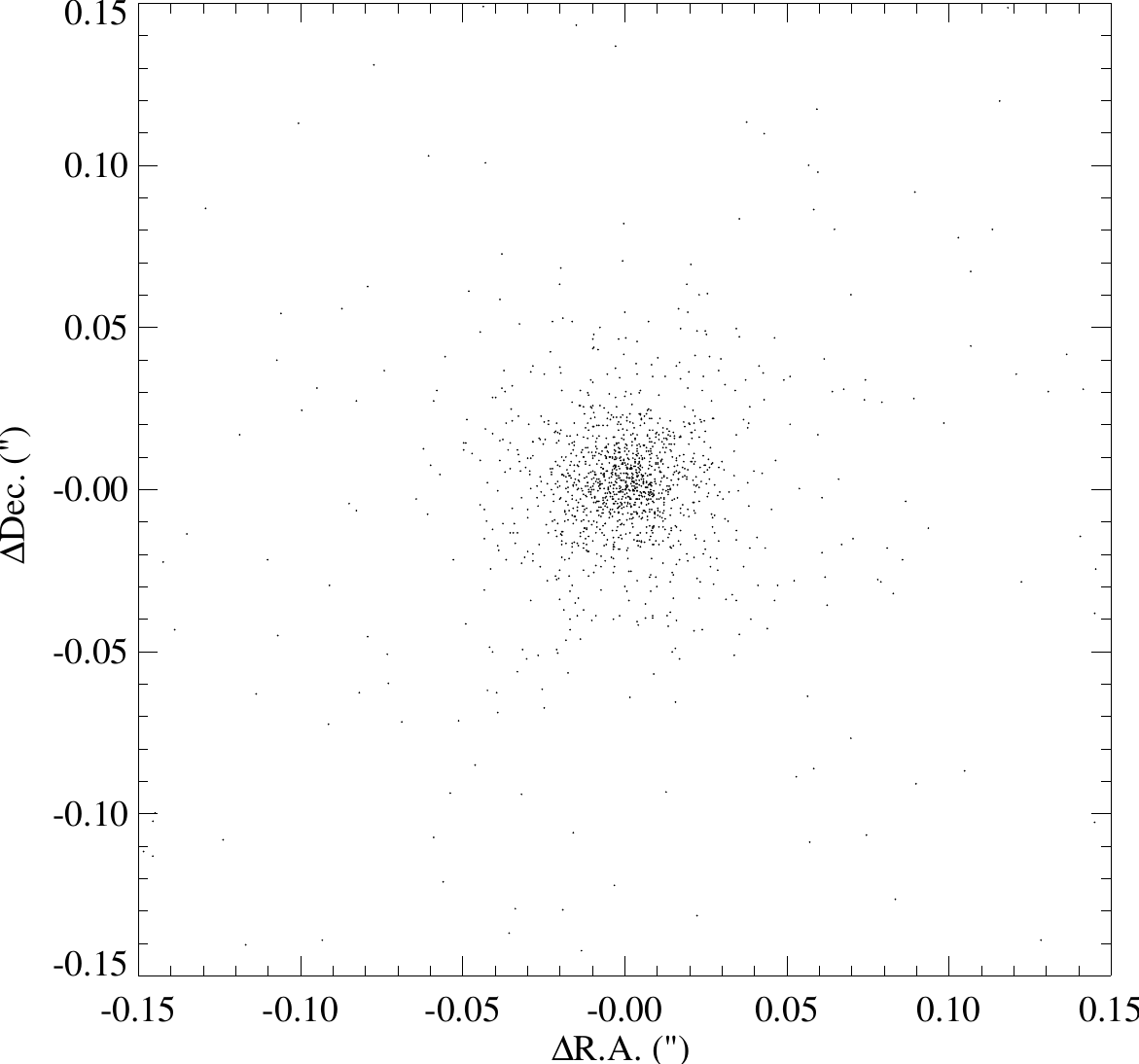}
\fi
\figcaption{\label{figure:astrom}%
Residual uncertainties in astrometry for a total of $\sim\,$1400 objects detected in the main UDF12 field, that were cross-correlated with sources from the original UDF catalog by
	Beckwith et al. (2006),
which was used for the astrometric registration for this project. Each point on the plot represents the remaining positional offsets between the sources, after having corrected for all the global $\lesssim\,$0\farcs005, after accounting for the positional uncertainty of each of the sources. As a result, the astrometry is sufficiently accurate to allow all the exposures in the dataset to be reliably combined, with no significant residuals remaining.}
\end{center}
\ifsubmodeapjs
  \end{figure*}
  \begin{figure*}[h]
\fi
\begin{center}
\ifsubmodeapjs
  \includegraphics[width=5in]{udf_matched_cat_radec_vector.eps}
\fi
\ifsubmodeastroph
  \includegraphics[width=5in]{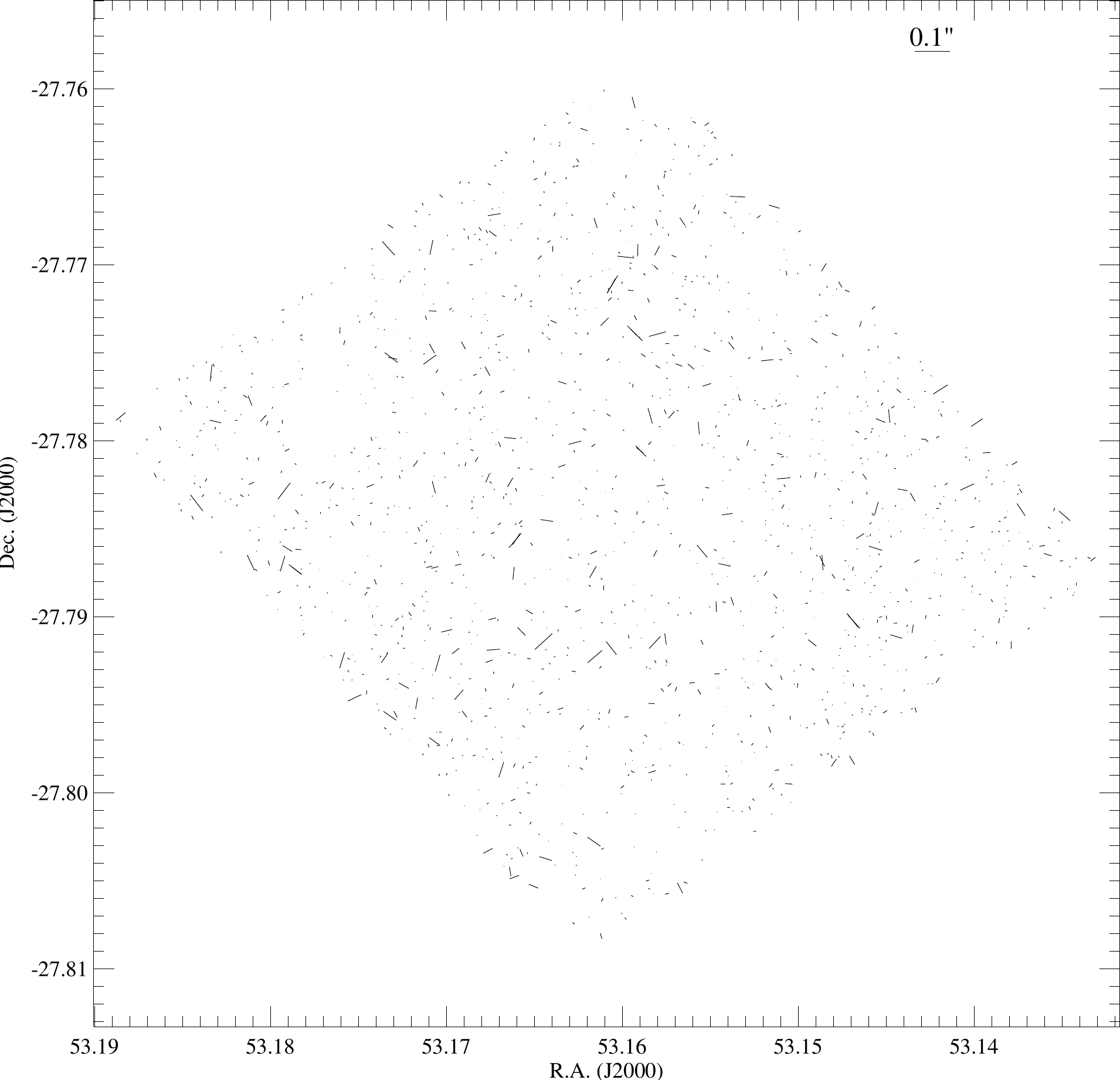}
\fi
\figcaption{\label{figure:vector}%
Residual uncertainties in astrometry represented as vectors on the UDF field, for each of the individual UDF12 sources that were matched to the original UDF catalog by
	Beckwith et al. (2006).
This indicates that there are no regions of net residual astrometric shift remaining across the image, thereby further confirming that the accuracy of the overall alignment is robust to the level of $\lesssim\,$0\farcs005, and sufficient to enable reliable combination of the full set of input exposures.}
\end{center}
\end{figure*}

The results of this procedure are shown in Figures~\ref{figure:astrom} and~\ref{figure:vector}, for the ACS/WFC and WFC3/IR images. Figure~\ref{figure:astrom} shows the distribution in right ascension and declination offsets between the reference catalog positions of the sources and those measured on the new data, after having solved for the astrometry as described above. Figure~\ref{figure:vector} is based on the same data as in Figure~\ref{figure:astrom}, but this time showing the residuals as a function of position across the field, where each vector indicates the mean residual in a grid of cells, each 40\arcsec on a side, where the size is chosen for display purposes to ensure a sufficient number of objects per cell while also providing a sufficient number of cells across the mosaic to show the general structure of the residuals. The residuals are generally below $\lesssim\,5$\,mas for the images, which is comparable to the typical on-orbit jitter of \HST\ and sufficient to mitigate all registration uncertainties.

\section{Final Data Products}\label{section:data}

\subsection{Mosaicdrizzle Combination}

We combined all the individual exposures, for each filter, into a final mosaics using ``inverse variance weighting'', whereby a weight map is created for each exposure, containing all the uncertainties that are intrinsic to a particular exposure (such as dark current and read-noise, and background sky noise, modified by the gain of the detectors as well as the flat field). Note that this excludes the additional Poisson terms from sources in the image, which can be added separately after the fact, if needed.

As an iterative step in the combination, we also removed remaining low-level background emission in each exposure from a masked version of the exposure, which was constructed from the full-depth image obtained by combining all the individual filters into a single image. This image was subsequently smoothed using both a small-scale 2-pixel Gaussian, as well as a larger-scale 10-pixel Gaussian, which enabled us to construct a mask of all the sources on both large scales (to exclude faint outer wings) as well as small scales (to exclude small remaining sources, which would otherwise impact the overall sky level determination).

The pixel scale for the output mosaics is driven by the detector plate scale and pixel size, together with the full width at half-maximum intensity (FWHM) of the PSF produced by the telescope optics. At the wavelengths of the WFC3/IR F105W to F160W observations, the {\it HST} PSF has a FWHM $\sim\,$0\farcs12$\,-\,$0\farcs18, which is subsequently convolved by the 0\farcs128 WFC3/IR detector pixel scale. Hence, the best PSF that could be recovered (without deconvolution), even in the ideal scenario of combining images using interlacing, which would minimize additional convolutions, still has a FWHM $\sim\,$0\farcs17$\,-\,$0\farcs19 in the final images. We choose an output pixel scale of 0\farcs06 pixel$^{-1}$ for the final WFC3/IR mosaics, providing adequate sampling of the PSF. We also chose a pixfrac parameter of 0.8, which is small enough to provide some reduction in the overall convolution as input pixels are mapped to the output plane, while at the same time not being too small, so that the overall pixel-to-pixel variation in the output weight map is not adversely affected. See 
	Koekemoer et al. (2002; 2011)
and
	Koekemoer et al (2011)
for further details about this parameter and its impacts on the final output images, in the context of deep imaging surveys with HST.

\begin{figure*}[t]
\begin{center}
\vspace{4cm}
\ifsubmodeapjs
  \includegraphics[width=7in]{udf.eps}
\fi
\ifsubmodeastroph
  \includegraphics[width=7in]{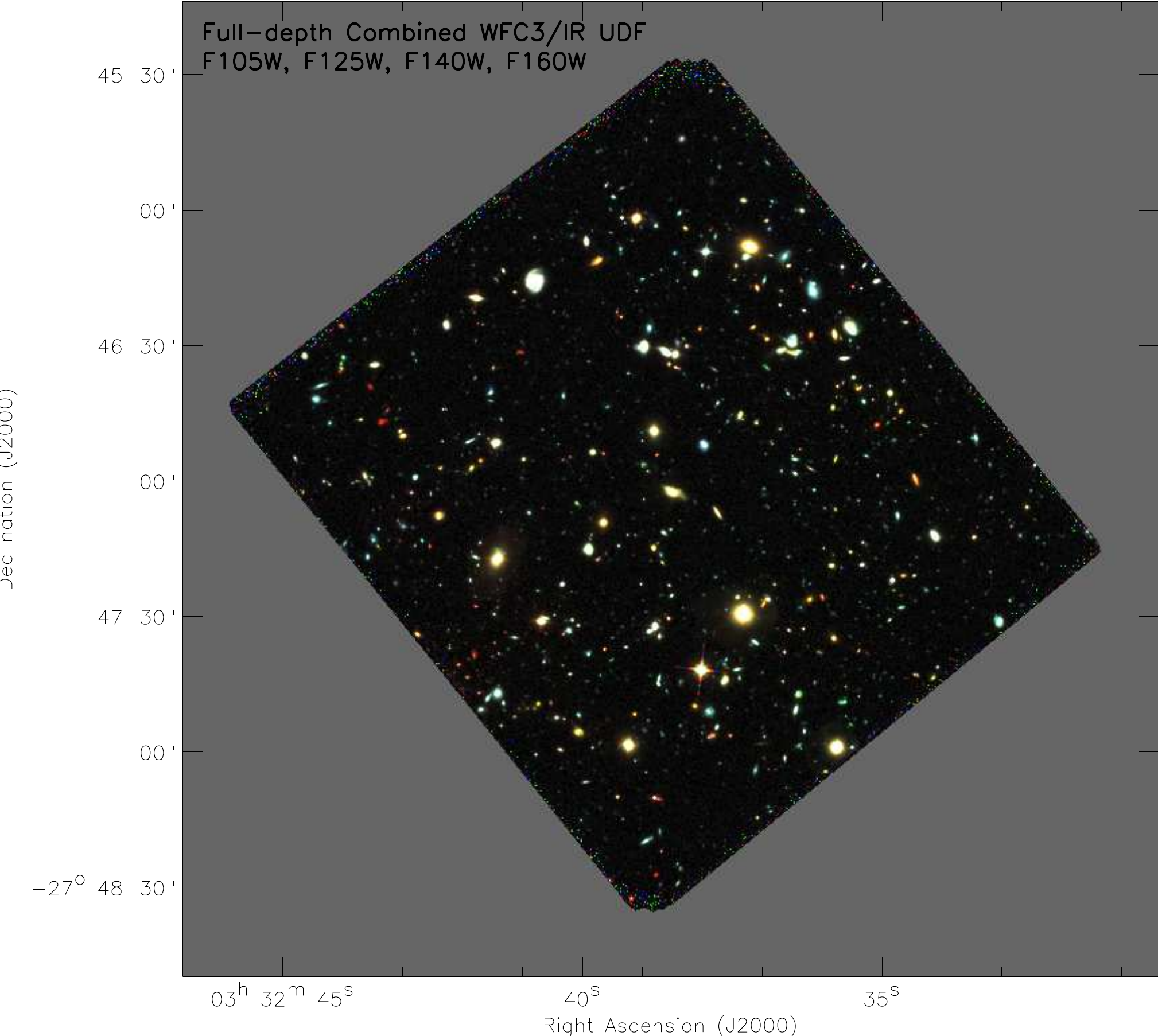}
\fi
\figcaption{\label{figure:udf}%
Image showing the full mosaics from our new WFC3/IR UDF12 data combined with the previous UDF09 and other data on this field, including all the WFC3/IR filters (F105W, F125W, F140W, F160W).}
\end{center}
\end{figure*}

Figure~\ref{figure:udf} shows the full combined mosaics obtained on the UDF main field. Three months after the completion of the observations, we release all the calibrated mosaics to the public via the STScI archive\footnote{http://archive.stsci.edu/prepds/hudf12/}, including the drizzled science mosaics as well as the inverse variance weight files that describe the noise associated with each pixel. Further updates on the project will be provided at the primary UDF12 project website\footnote{http://udf12.arizona.edu/} as needed.

\subsection{Photometric Limiting Depth Validation}\label{section:photom}

We have carried out a series of photometric and limiting depth tests on the full combined WFC3/IR mosaics, aiming to validate the depth achieved in absolute terms as well as relative to the previous data on this field. In order to quantify the limiting depth across the mosaic, we first constructed a full-depth, full-filter mosaic using all the WFC3/IR observations on the UDF, in all four filters (F105W, F125W, F140W, F160W). This broad-band image provided extremely deep sensitivity for masking out the extended faint wings of sources, which is necessary in order to obtain genuinely source-free regions in order for an accurate background r.m.s. estimate to be obtained. In order to improve the signal-to-noise with which faint objects were masked, we created smoothed versions of this full-depth image by convolving it with a small-scale Gaussian (2 pixels FWHM, aimed at detecting and masking all the faint sources), as well as a larger-scale Gaussian (10 pixels FWHM), which successfully masked all extended emission around larger sources.

\begin{figure*}[t]
\begin{center}
\vspace{4cm}
\ifsubmodeapjs
  \includegraphics[width=7in]{udf_main_wfc3_allfilters_udfonly_60mas_v3_6_drz.eps}
\fi
\ifsubmodeastroph
  \includegraphics[width=7in]{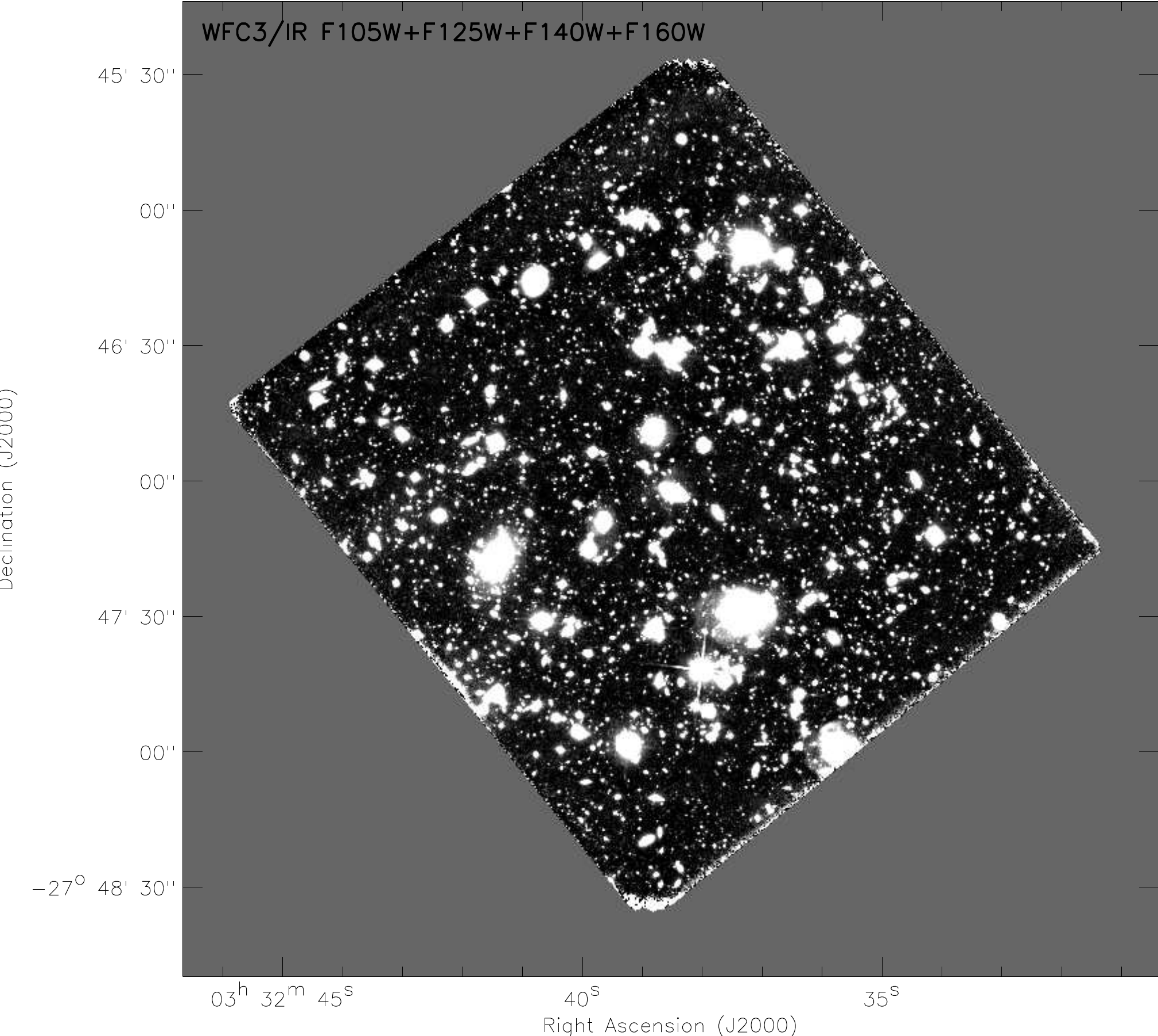}
\fi
\figcaption{\label{figure:fulldepth}%
Full-depth UDF image, created by combining all the F105W, F125W, F140W and F160W exposures, that was used to create the object mask subsequently used for the blank sky statistical measurements for the limiting depth calculations.}
\end{center}
\end{figure*}

In Figure~\ref{figure:fulldepth} we show the full-depth image obtained from all four filters, which was used to create the object mask that we subsequently applied. The object mask excludes about 45\% of all the pixels in the mosaic; the remaining pixels were then considered to represent the pure sky background (along with potentially exceedingly faint sources that are not included in the object mask). The statistics of these pixels were analyzed using several different tests, in order to determine the r.m.s. values on small scales as well as determining the global uniformity of r.m.s depth across the mosaic.  These pixel-to-pixel r.m.s. values, after accounting for correlated noise, then provide a direct estimate of the limiting sensitivity of the mosaic.

\begin{figure*}[t]
\begin{center}
\vspace{4cm}
\ifsubmodeapjs
  \includegraphics[width=7in]{udf_main_wfc3_f105w_full_60mas_v3_6_drz.eps}
\fi
\ifsubmodeastroph
  \includegraphics[width=7in]{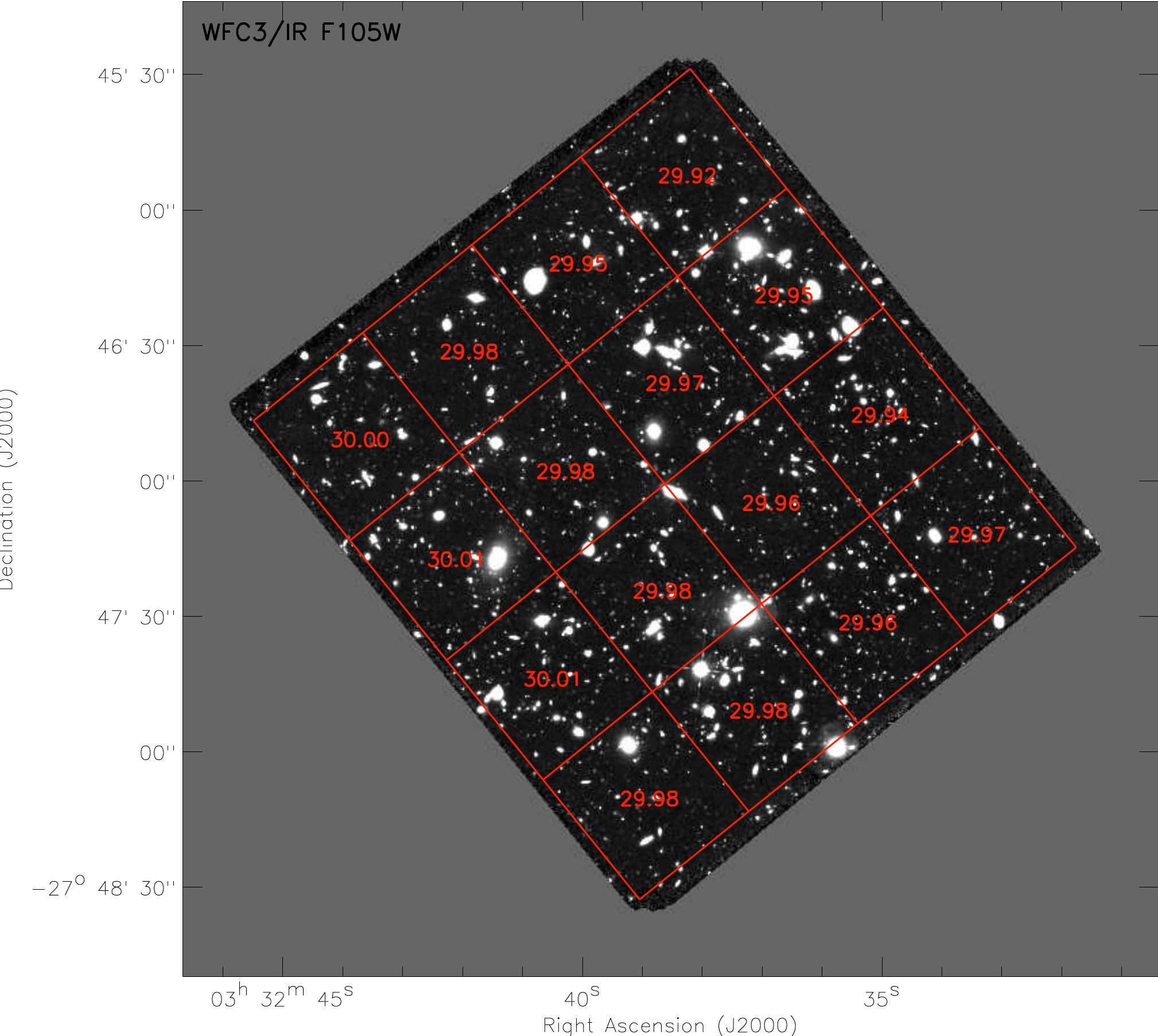}
\fi
\figcaption{\label{figure:depthtestsf105w}%
Measured 5$\sigma$ limiting sensitivity across the full-depth F105W UDF mosaic. See text for further details.}
\end{center}
\end{figure*}

\begin{figure*}[t]
\begin{center}
\vspace{4cm}
\ifsubmodeapjs
  \includegraphics[width=7in]{udf_main_wfc3_f160w_full_60mas_v3_6_drz.eps}
\fi
\ifsubmodeastroph
  \includegraphics[width=7in]{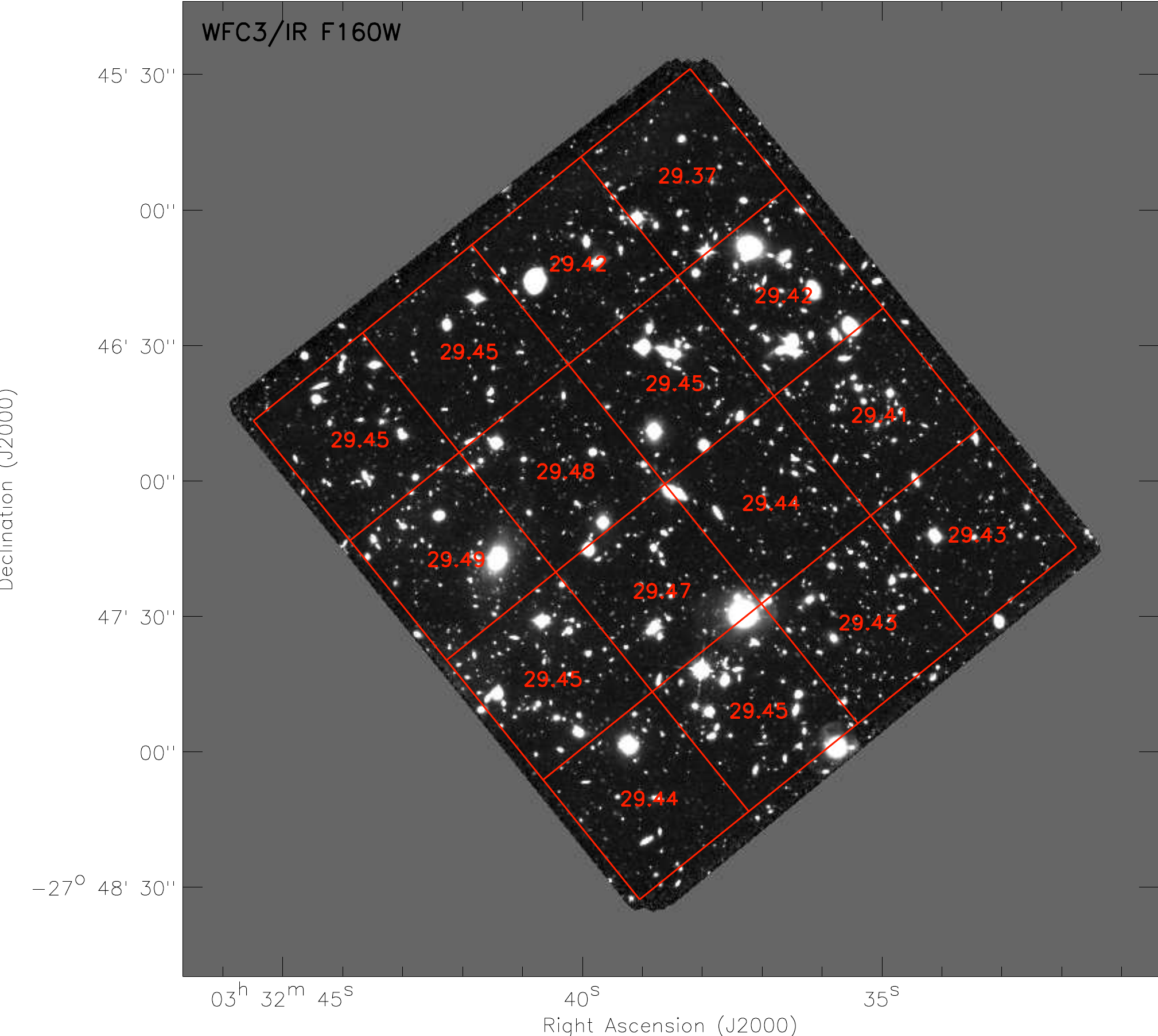}
\fi
\figcaption{\label{figure:depthtestsf160w}%
Measured 5$\sigma$ limiting sensitivity across the full-depth F160W UDF mosaic. See text for further details.}
\end{center}
\end{figure*}

The first test involved dividing the mosaic into a regular grid of cells, in order to determine the relative degree of depth variation across the mosaic, as well as the degree of flatness in the residual sky level. This test reveals the degree of uniformity in detection sensitivity between different cells, as well as the impact of any remaining large-scale residuals in the background sky, which would serve to broaden the global measured r.m.s. as compared with the average of all the individual r.m.s. measurements obtained in the different cells. The results from these tests are shown in Figures~\ref{figure:depthtestsf105w}$\,-\,$\ref{figure:depthtestsf160w}, where we translate the pixel-to-pixel r.m.s. into 5$\sigma$ limiting magnitudes in apertures of diameter 0$\farcs$4, 0$\farcs$44, 0$\farcs$47 and 0$\farcs$50 respectively for F105W, F125W, F140W and F160W (corresponding to 70\% of the total enclosed flux), after having accounted also for the presence of correlated noise by comparing the science images with the inverse variance weight maps.

In addition, the limiting depths are relatively uniform across the field, with residual variations $\lesssim\,$0.03 magnitude except for occasional areas of somewhat lower weight that correspond to known large defects on the detector. Finally, the global measured r.m.s. values from these tests agree very well with the average r.m.s. from the individual cells ($\lesssim\,$0.01 magnitudes), indicating that the background residual sky structure is globally flat with no significant impact on the overall r.m.s. These results are presented in Table~\ref{table:fulldepth}.

We also performed blank-aperture tests, calculating the 5$\sigma$ depths based on a total of 15,000 apertures placed within blank sky regions across the mosaic. These tests include fitting only the negative half of the pixel distribution, in order to exclude low-level positive sources, and used the same sized apertures as the previous tests. These results are also presented in Table~\ref{table:fulldepth}. The results agree very well, indicating that the depths achieved match the expected sensitivities for each of the filters, for these full-depth mosaics, demonstrating that we achieve our proposed limiting depths of AB$\,\sim\,$30 for F105W, and AB$\,\sim\,$29.5 for F125W, F140W and F160W.

\begin{deluxetable}{lccc}
\tablecaption{\label{table:fulldepth}
	Final limiting 5$\sigma$ sensitivities\tablenotemark{a}}
\tablehead{%
Filter		& Negative Gaussian
				& Average r.m.s.
						& Global r.m.s.}
\startdata
F105W		& 29.98		& 29.97		& 29.97		\\
F125W		& 29.55		& 29.53		& 29.52		\\
F140W		& 29.51		& 29.49		& 29.48		\\
F160W		& 29.46		& 29.45		& 29.45
\enddata
\tablenotetext{a}{All measured depths are for our final, full combined mosaics, representing the 5$\sigma$ limiting depth in apertures of diameter 0$\farcs$4, 0$\farcs$44, 0$\farcs$47 and 0$\farcs$50 respectively for F105W, F125W, F140W and F160W.\hfill\break}
\end{deluxetable}

\section{Summary}\label{section:summary}

We have described the 2012 Hubble Ultra Deep Field campaign (UDF12), a large 128-orbit Cycle 19 \HST\ program aimed at extending previous WFC3/IR observations of the UDF by quadrupling the exposure time in the F105W filter, adding a completely new F140W filter, and extending the F160W exposure time by 50\%. The project is aimed at determining the role played by galaxies in reionizing the universe, and includes obtaining a robust determination of the star formation density at $z$$\,\gtrsim\,$8, improving measurements of the ultraviolet continuum slope at $z$$\,\sim\,$7$\,-\,$8, facilitating the construction of new samples of $z$$\,\sim\,$9$\,-\,$10 candidates, and enabling the detection of sources up to $z$$\,\sim\,$12. For this project we committed to combining these and other WFC3/IR imaging observations of the UDF area into a single homogeneous dataset, to provide the deepest near-infrared observations of the sky currently achievable. We have described the observational aspects of the survey as motivated by its scientific goals, and have presented a detailed description of the data reduction procedures and products from the survey. We release the full combined mosaics, comprising a single, unified set of mosaics of the UDF, providing the deepest near-infrared blank-field view obtained of the universe to date, reaching magnitudes as deep as AB$\,\sim\,$30 in the near-infrared, and yielding a legacy dataset on this field of lasting scientific value to the community.

\acknowledgments

We wish to thank the Hubble observing team as well as our Program Coordinator, Shelley Meyett, for help in scheduling and executing this program. Support for {\it HST} Program GO-12498 is provided by NASA through a grant from the Space Telescope Science Institute, which is operated by the Association of Universities for Research in Astronomy, Incorporated, under NASA contract NAS5-26555. JSD, RAAB, PRD, TAT and  VW acknowledge the support of the European Research Council through the award of an Advanced Grant to JSD. JSD and RJM also acknowledge the support of the Royal Society via a Wolfson Research Merit Award and a University Research Fellowship respectively. ABR and EFCL acknowledge the support of the UK Science and Technology Facilities Council. VW acknowledge the support of the European Research Council through the award of a Starting Grant to VW. SC acknowledges the support of the European Commission through the Marie Curie Initial Training Network ELIXIR.

Facilities: \facility{HST(ACS,WFC3)}

\bibliographystyle{apj}
\bibliography{apjmnemonic,manuscript}

\clearpage

\begin{figure*}[t]
\begin{center}
\vspace{4cm}
\ifsubmodeapjs
  \includegraphics[width=7in]{udf_main_wfc3_f125w_full_60mas_v3_6_drz.eps}
\fi
\ifsubmodeastroph
  \includegraphics[width=7in]{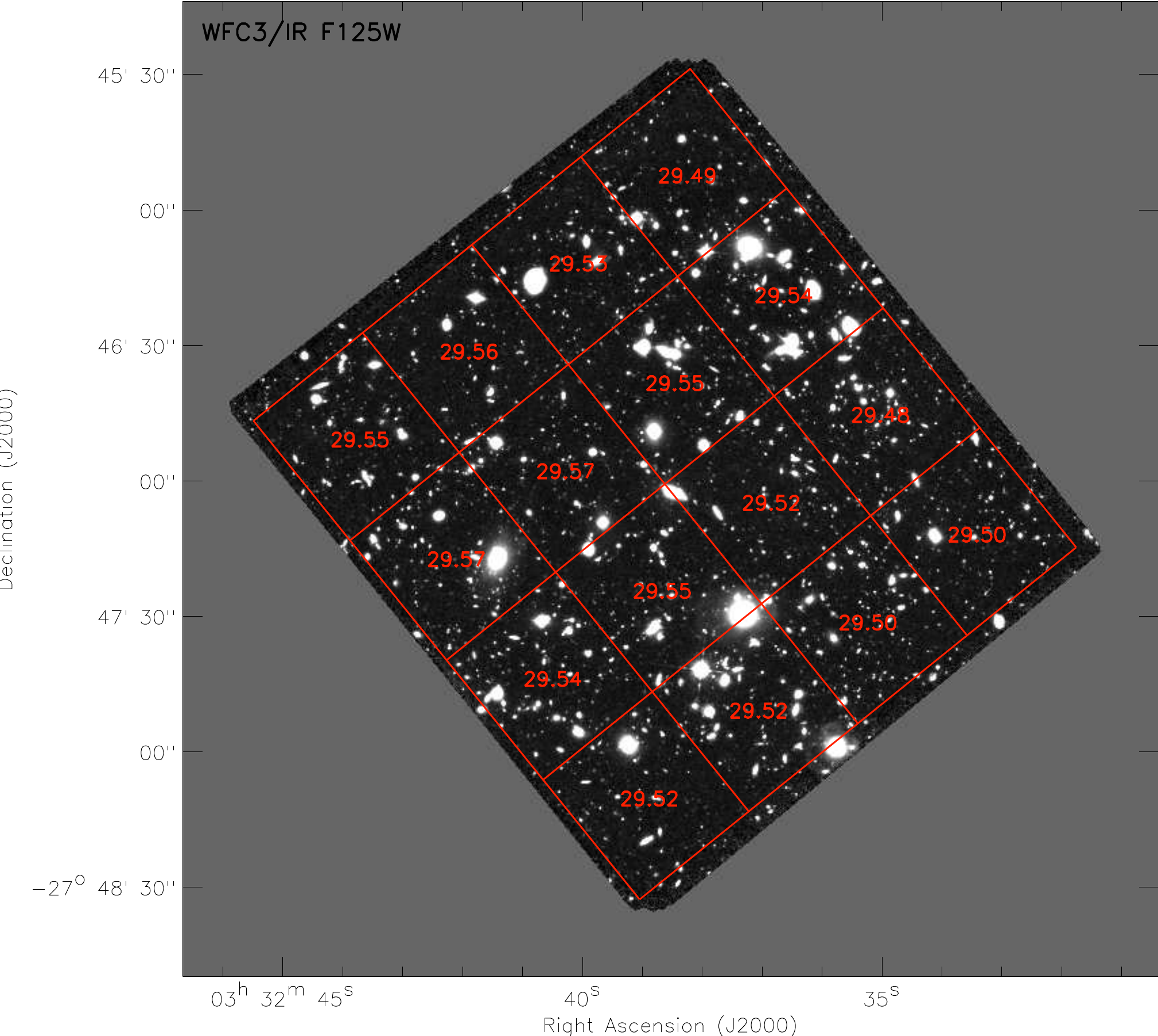}
\fi
\figcaption{\label{figure:depthtestsf125w}%
Measured 5$\sigma$ limiting sensitivity across the full-depth F125W UDF mosaic. See text for further details.}
\end{center}
\end{figure*}

\begin{figure*}[t]
\begin{center}
\vspace{4cm}
\ifsubmodeapjs
  \includegraphics[width=7in]{udf_main_wfc3_f140w_full_60mas_v3_6_drz.eps}
\fi
\ifsubmodeastroph
  \includegraphics[width=7in]{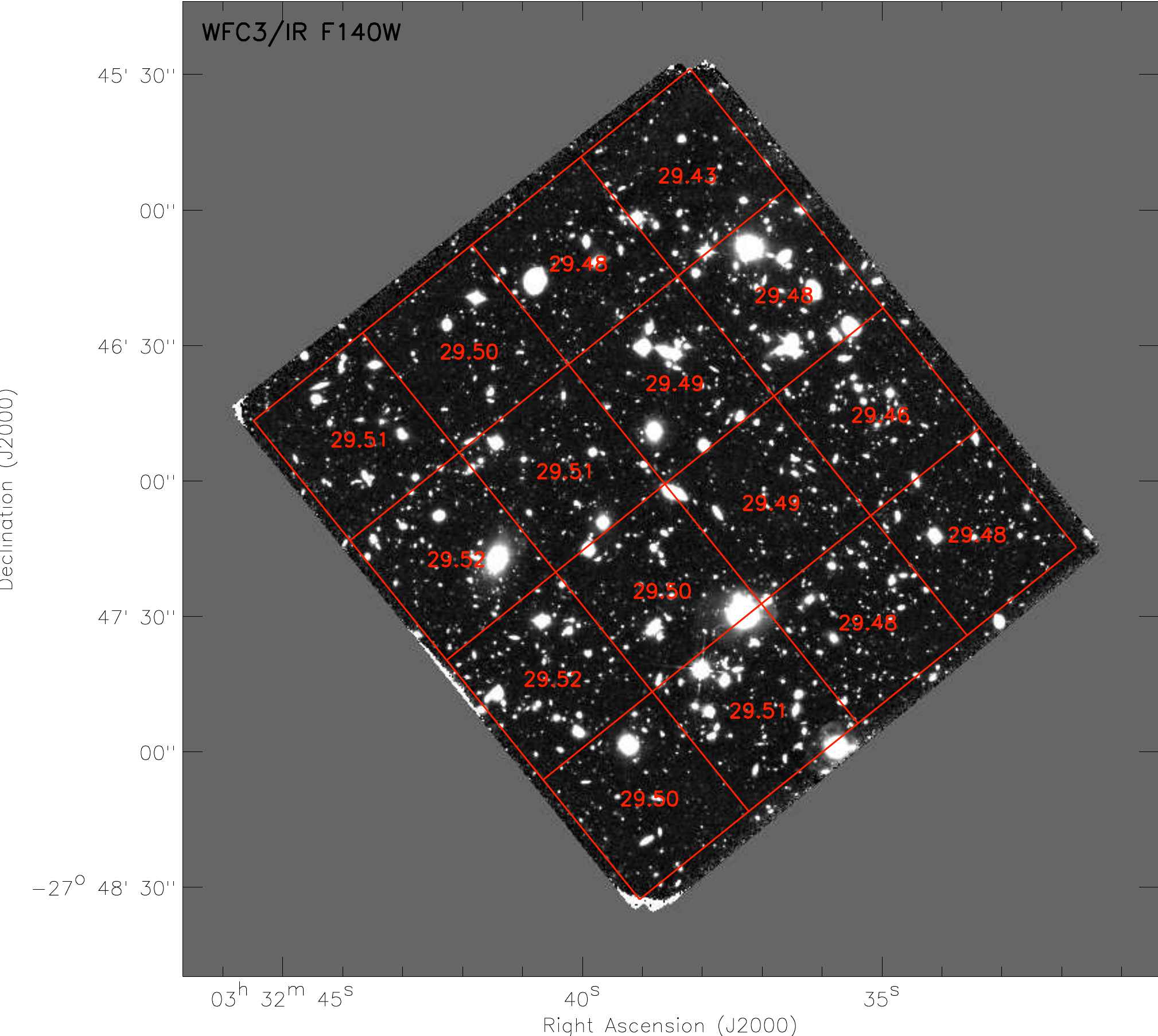}
\fi
\figcaption{\label{figure:depthtestsf140w}%
Measured 5$\sigma$ limiting sensitivity across the full-depth F140W UDF mosaic. See text for further details.}
\end{center}
\end{figure*}

\end{document}